# Self-organised criticality and *1/f* noise in single-channel current of voltage-dependent anion channel

Jyotirmoy Banerjee[1], Mahendra. K. Verma[2], Smarajit Manna[1] & Subhendu Ghosh[3,1*]

1. Department of Biophysics
University of Delhi South Campus
Benito Juarez Road
New Delhi 110021, India

2. Department of Physics
Indian Institute of Technology
Kanpur 208016, India

3. Department of Animal Sciences, School of Life Sciences,
University of Hyderabad,
Hyderabad 500046, India

* Corresponding Author:
Department of Animal Sciences, School of Life Sciences,
University of Hyderabad,
Hyderabad 500046, India.
e-mail: sgsl@uohyd.ernet.in

PACS. 87.15.-v: Biomolecules: structure and physical properties.

PACS. 87.15.Aa: Theory and modeling; computer simulation.

PACS. 87.15.Ya: Fluctuations

**Running title:** Noise analysis of VDAC.

## Abstract:

Noise profile of Voltage Dependent Anion Channel (VDAC) is investigated in open channel state. Single-channel currents through VDAC from mitochondria of rat brain reconstituted into a planar lipid bilayer are recorded under different voltage clamped conditions across the membrane. Power Spectrum analysis of current indicates powerlaw noise of *1/f* nature. Moreover, this *1/f* nature of the open channel noise is seen throughout the range of applied membrane potential –30 to +30 mV. It is being proposed that *1/f* noise in open ion channel arises out of obstruction in the passage of ions across the membrane. The process is recognised as a phenomenon of self-organized-criticality (SOC) like sandpile avalanche and other physical systems. Based on SOC it has been theoretically established that the system of ion channel follows powerlaw noise as observed in our experiments. We also show that the first-time return probability of current fluctuations obeys a powerlaw distribution.

Fluctuation analysis has been an important aspect of research in various disciplines e.g. physics, chemistry, geology, environmental sciences, biology and others. In physiology, a good number of reports related to noise have come up during recent years [1]. Advanced electrophysiological experiments, as reported earlier, enable us to record currents through a single or a group of channels on a cell or a lipid bilayer membrane under voltage clamped conditions [2]. Recent development on quantification of noise at the ion channel level has thrown light on the phenomenon of transport of ions and metabolites across cell membrane and its mechanisms. Particularly, neuronal communications and transfer of action potential between a pair of neurons through synapses has been recognised to be the key parameter for functioning of the brain. Synaptic noise, a kind of channel noise, plays an important role in this process [1]. In this paper we discuss the time series behaviour of the channel currents and the associated noise. Rostovtseva and Bezrukov have looked into the noise in VDAC during ATP transport [3] and in OmpF during PEG transport [4], and reported white noise.

In the present work we report the properties of noise of Voltage Dependent Anion Channel (VDAC), and construct a theoretical model to explain the noise pattern. VDAC is an abundant protein in the outer mitochondrial membrane, which forms large voltage gated pore in planar lipid bilayers, and act as the pathway for the movement of substances in and out of the mitochondria by passive diffusion [4, 5]. VDAC essentially plays an important role in the transport of ATP, ions, and other metabolites between the mitochondrion and the cytoplasm. All VDACs form channel with roughly similar single channel conductance (4.1 nS to 4.5 nS, in 1 M KCl). This voltage gated pore has an effective diameter 2.5 nm to 3 nm and develops cation preference in closed or lower conductance states [7, 8, 9, 10, 11]. In our previous papers, the gating behaviour of VDAC has been discussed along with the existence of its capacitance [12, 13].

Bezrukov and Winterhalter [14] observed *1/f* noise for similar porin channels. There are reports on experimental evidence and analysis of open channel noise by Mak and Webb in alamethicin [15], by Hainsworth *et al.* in $K^+$ channels from sarcoplasmic reticulum [16], and by Zhou *et al.* in CFTR channel pore [17]. Theoretical understanding of *1/f* noise in ion-channels is in its infancy. Earlier, Bezrukov and Winterhalter [14] argued that *1/f* noise in ion-channel current is not a fundamental property of nonequilibrium transport phenomena, rather it reflects the complex hierarchy of equilibrium protein dynamics. Siwy *et al.* (2002) have discussed the aforesaid problem in biological and synthetic channels on lipid bilayer membrane and claimed that the powerlaw noise (non-stationary) originates from the channel's opening and closing [18]. They analysed the channel gating in view of Markov process and tested through Smoluchowski-Chapman-Kolmogorov (SCK) equation. It has been reported that the maltodextrin sugar molecules passes through the maltoporin channel with a high collisional friction [19] which might be the cause of the *1/f* noise in current.

Generically, systems under equilibrium have thermal noise, which are typically uncorrelated or white. As mentioned earlier, Bezrukov [14] and others have reported that noise spectrum of ion-channel is *1/f*. Hence noise in ion-channel is non-thermal, and therefore, correlated. Also, the transport process through ion-channel is non-equilibrium or driven because the current is directed due to external potential.

In this paper we attempt to provide a theoretical model for ion-channel noise. Earlier, scientists have proposed various schemes to explain *1/f* noise. Some of them are activated random processes, diffusion [20], self-organized criticality (SOC), etc. The non-equilibrium nature of ion-channels rules out diffusion. The transport in ion-channel is a steady-state process, and it cannot be modeled satisfactorily by relaxation process. Hence, activated-random process appears to be ruled out as well. Considering that the transport in ion-channel is a non-equilibrium steady-state flow, self-organized criticality (SOC) appears to be the most promising candidate to explain *1/f* noise in ion-channel [21]. That is the reason why we focused on SOC in this paper.

Bak *et al.* proposed self-organised critically as the source of *1/f* noise in sandpile avalanches [22, 23]. They illustrated their theory using random avalanches in sand-pile as events. Sand is poured slowly at random position of the sandpile. At critical conditions, avalanches of various sizes and duration occur in the system. In the present work we investigate and analyse the open channel (single) noise of VDAC in lipid bilayer membrane at different potentials applied across the membrane. We claim that the random obstruction of ions during the passage through an ion channel follows SOC dynamics.

The experimental set up is the same as that of Bezrukov and Winterhalter [14] and Banerjee and Ghosh [24]. VDAC was purified from rat brain mitochondria using the method of De Pinto *et al*. [25], and reconstituted into the planar lipid bilayers according to the method of Roos *et al*. [26]. Aqueous compartments on both sides of the bilayer membrane are connected to an integrating patch amplifier Axopatch 200 A (Axon Instruments, USA) through a matched pair of Ag/AgCl electrodes. Axopatch 200A was connected to an IBM computer through an interface Digidata 1322A (Axon Instruments, USA). Channel current due to transport of $K^+$ and $Cl^-$ was recorded using the data acquisition software Clampex (pClamp 9.0, Axon Instruments, USA). The experiment was performed on anti-vibration table (TMC, USA) to avoid any vibrational noise. Single channel recording of VDAC was performed in a symmetric bath solution.

We record ion current arising from single-channel in presence of externally applied potential across the membrane. A time series consists of open and closed states. Here, we focus only on open states, whose typical time trace for +25mV is shown in Fig. 1. Clearly, the amplitude of noise has a large variation with sufficient number of large bursts. We compute the following noise characteristics: (a) probability distribution of large current amplitudes (*P(I)* vs. *I*), (b) probability distribution function of intervals between two consecutive current bursts, and (c) power spectral density *S(f)*.

To compute the probability distribution of large current amplitudes *P(I)*, we take time traces for open states with voltage of -25mV, and obtain dataset of 241764 points. Now we plot a histogram of *P(I)* vs. *I* in log-log scale, which is shown in Fig. 2. We find that $P(I) \alpha I^{-a}$ with $a = 5.0 \pm 0.28$. Hence, the amplitudes of large fluctuations obey a power law. This result is similar to the avalanche size or earthquake size distribution, and it clearly indicates that the transport in the ion-channel is a nonequilibrium process. It may be noted that the equilibrium processes have typically Gaussian probability distribution.

Now we compute the statistics of time-intervals between two successive large current fluctuations in current traces at a constant volatge (+25mV and +20mV). We take single time trace and compute the standard deviation of the fluctuations. The large current fluctuations are marked by taking noise signals which are three times the standard-deviation. After the large current fluctuations have been identified, we measure the time-interval $\tau$ between two consecutive large current fluctuations. A histogram (of bin size 20) of $\tau$ has been plotted in Figs. 3a and 3b corresponding to volatges +25mV and +20mV respectively. We find that $P(\tau) \sim \tau^{-\beta}$ with $\beta = 1.9 \pm 0.1$ at +25 mV and $1.8 \pm 0.2$ for +20 mV +20 mV. Similar analysis for different current cutoffs are performed. We find that the level of $P(\tau)$ changes with cutoff, but the slopes are approximately the same (within error bar). We also notice that the single channel traces with low channel conductance and closed states do not follow any powerlaw.

Computation of power spectral density $S(f)$ was done using the software Clampfit. We used 2064-point vectors. The plot of spectral density $S(f)$ versus frequency $f$ for a current time trace of open VDAC channel at +25mV is shown in Fig. 4. It is evident from the figure that powerlaw $f^{\alpha}$ fits quite well to $S(f)$ vs. $f$ plot for frequency range of more than two decades. Table 1 shows the slope ($\alpha$) of power spectrum of single channel traces of VDAC at various applied voltages. The slope ranges from 0.72 to 1.05. These results indicate that the noise exhibits *1/f* spectrum which is consistent with that of Bezrukov and Winterhalter [3]. This is maintained throughout the range of potential $\pm$ 30mV.

The slope of power spectrum has no systematic variation with applied voltage, and the values of the slope are reasonably close to 1.0. The power spectrum is not reproducible because protein channels get embedded at different locations in the bilayer membrane in different experiments with a little variation in conformation. Also, the channels have intrinsic properties which must affect the noise characteristics.

The positivity of spectral index ($\alpha > 0$) implies that noise is correlated at large-scales, and correlation lengths are large. The divergence of correlation length is referred to as critical behaviour for the system. Interestingly, ion channels exhibit powerlaw behaviour for the noise under steady state without any fine tuning. Hence the above behaviour is referred to as self-organized criticality. In a popular sandpile model of Bak et al. [22], when the pile is large and the pouring rate is small, the activity is described by random linear superposition of individual avalanche signals. Bak *et al.* [22] derived the power spectrum and found the spectral index to be close to 1.0. Many researchers challenged Bak *et al.*'s assumption that avalanches are uncorrelated. When the input rate is significant, avalanches overlap, and they are correlated. Among others, Maslov, Paczuski, and Bak [27], and Hwa and Kardar [28] studied various systems under these limits, and found the signals to be correlated. They however found the power spectrum to be again a powerlaw at low frequencies.

In the following, we propose a novel mechanism to explain *1/f* noise in ion channel. Our derivation is inspired by a model for intermittency route to chaos [29]. Recently some interesting studies have been reported which show that the waiting-time or the first-time return probability between earthquakes of magnitude M or greater has a powerlaw distribution, i.e., $P(\tau) \sim \tau^{-\beta}$, where $\tau$ is the time gap between two earthquakes of magnitude M or greater and $\beta$ is the exponent [30, 31]. The correlation is built up by the accumulation of stresses over time.

To argue for the self-organized-criticality of the VDAC channel dynamics it was shown that the first-time return probability of large fluctuation of current in the time series exhibit powerlaw distribution (Figs. 3a and 3b), i.e., the probability of occurence of large (three times the standard deviation or greater) magnitude of current flutuations obeys a powerlaw, i.e.,

$$P(\tau) \sim \tau^{-\beta} \quad [1]$$

It is reasonable to expect that dynamical variation of pore-width can induce correlations because ions pile up near the barrier. The large fluctuations arise because the ions pass through the pore intermittently. So depending on the size of the ion pile, large current fluctuations occur at a very short interval of time as well as long interval. The above-mentioned powerlaw distribution for the first-time return may be due to variable width. It may be noted that the shape of ion channel is highly dynamic; the irregular inner surface of the channel act as barriers.

Now imagine that a surge of current occurs at time $t_i$ due to removal of blockage with $t_i$ being random. The current flow is modelled by *a(t)* as given below:

$$a(t) = -A \sum \delta(t - t_i) \quad [2]$$

where A is the amplitude of the noise, which is taken to be the same for all time as an approximation. The quantity $\tau_i = t_{i+1} - t_i$ is correlated. The correlation function is defined as

$$C(\tau) = \langle a(t)a(t+\tau) \rangle \quad [3]$$

We can choose $t = 0$ under the assumption that process is stationary in time. We use the definition of correlation function as conditional probability of finding an event at time $\tau$ given that an event took place at time 0. For convenience, we use discrete time with interval $\Delta$. If $\tau = n\Delta$, then

$$C(\Delta) = P(\Delta)$$
$$C(2\Delta) = C(\Delta)P(\Delta) + P(2\Delta)$$
$$\vdots$$
$$C(n\Delta) = \sum_{k=0}^{n-1} C[(n-k)\Delta]P(k\Delta) + \delta_{n,0}$$

where $P(\tau)$ has been defined in Eq. (1). We take $P(0) = 0$ and $C(0) = 1$. Fourier transform of the last equation yields

$$S(f) = \frac{1}{1 - \hat{P}(f)} \quad \text{...........[4]}$$

where $\hat{P}(f)$ is the Fourier transform of $P(\tau)$. Fourier transform of $P(\tau) \propto \tau^{-\beta}$ yields $\hat{P}(f) = 1 - Bf^{\beta-1}$. Therefore,

$$S(f) = Bf^{1-\beta} \quad \text{.............[5]}$$

which is $1/f^\alpha$ noise with $\beta = 1 + \alpha$.

The values of α and β as mentioned in the experimental results are consistent with the realtion β = 1 + α. It may be noted that the above relationship is based on an assumption that A is constant, which is not valid in general situations. A generalization of the above calculation is underway for different distribution of A. Here we focus on large events which occur after large time interval (small frequencies). A technical point is in order: Sanchez et al. [32] argue that SOC systems exhibit exponential waiting-time distribution, while "turbulent" systems exhibit power-law waiting-time distribution. However, here we do not differentiate "turbulent" and SOC mechanisms, which have many similarities.

We would like to contrast our theoretical result with white noise reported by Rostovtseva and Bezrukov [3]. They studied ATP transport through VDAC of *Neurospora crassa*, and found the power spectra to be proportional to $f^0$ (white noise). This implies that current during ATP transport must be uncorrelated ($\langle a(t)a(t+\tau)\rangle = A\delta(\tau)$), in contrast to small ion (e.g., K$^+$, Cl$^-$) transport which is correlated. The difference is probably due to the large size of the ATP molecules (several nanometers, comparable to the pore size). Because of large size, a single ATP molecule crosses the pore at a time, therefore crossings will be uncorrelated. We can draw the following analogy: ATP transport is like people coming into a room through a narrow door, while the open ion flow is like a canal whose flow is controlled by a gate in a dam. We also point out that the mechanism of noise production by friction as proposed in reference [19] has certain similarities with our proposal based on blockages in ion-channel.

*1/f* noise has been reported in physical systems like earthquakes [30, 31, 32], surface growth under quenched disorder [33], invasion percolation [34], biological systems like evolution [36] and cognition [35], time series taken from electronic currents, networks, market, and hosts of other systems. Current understanding is that *1/f* noise is not caused by any unique scheme, but different schemes may be at play in different systems. SOC however is one of the popular mechanism which appears to be applicable in many of the above systems.

In conclusion, the noisy patterns in the current fluctuations in open VDAC channel is due to random but correlated obstruction of ions during the passage through the channel. Powerlaw distribution indicates the phenomenon of self-organized criticality. This self-organized criticality, we propose, is the cause of *1/f* noise in open state VDAC.

**Acknowledgements:** The theoretical ideas discussed in this paper were born during discussion of MKV with Mustansir Barma and Supriya Krishnamurthy many years ago. MKV thanks them as well as Sutapa Mukherjee, Amit Dutta, S. A. Ramakrishna for useful discussions.

**Tables**

| Applied Voltage (mV) | Slope ($\alpha$) |
|---|---|
| +15 | 1.05 |
| +20 | 0.72 |
| +25 | 0.92 |
| +30 | 0.89 |
| -15 | 0.88 |
| -20 | 0.82 |
| -25 | 0.85 |
| -30 | 0.98 |

Table 1:The spectral indices of power spectrum $S(f)=1/f^{\alpha}$ of single channel current traces of VDAC at various applied voltages.

# Figure Legends

Fig. 1. Continuous current trace (open state) of rat brain VDAC at +25mV. Membrane bathing solution consisted of 500mM KCl, 10mM Hepes, and 5mM $MgCl_2$. Its pH value was 7.4. The experiment was done at room temperature (23-25°C).

Fig. 2: Log-log plot of probability distribution of magnitude of current fluctuation $P(I)$ vs. $I$ at -25 mV. The data size is 24176 points.

Figure 3(a,b). The log-log plot of first-time return probability of large current fluctuations (three standard deviation or higher) $P(\tau)$ vs. $\tau$ at bin size 20ms for the current traces at (a) +25 mV (b) +20mV. The fits are linear in log-log plot with slopes -1.9 ± 0.1 and -1.8 ± 0.2 respectively. The data size is 2064 points.

Fig. 4. The plot of power spectral density $S(f)$ vs. frequency $f$ for a full open state at +25 mV. A powerlaw $1/f^\alpha$, $\alpha$ = -0.92, fits the data quite well.

# Figures

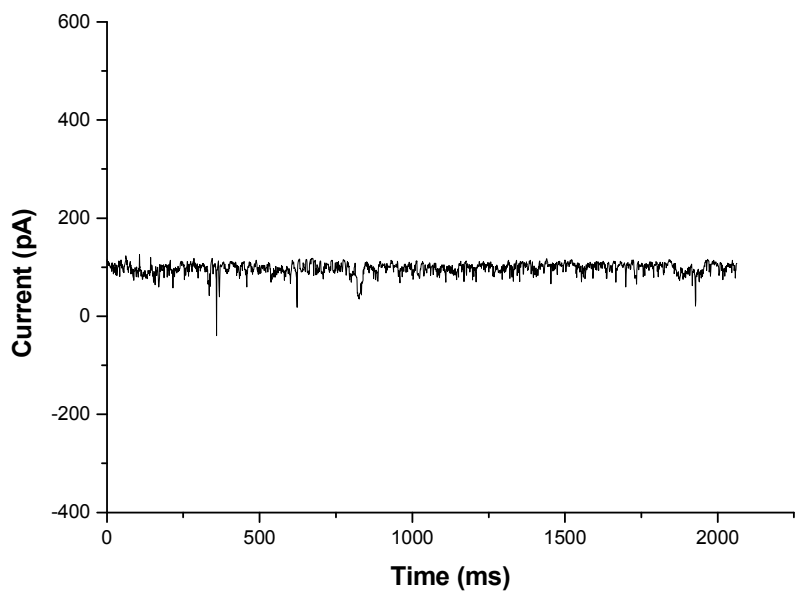

Figure 1.

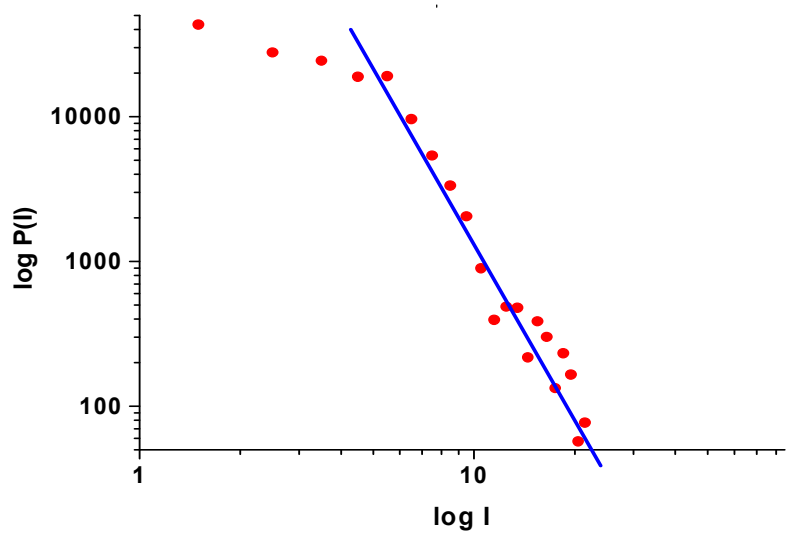

Figure. 2

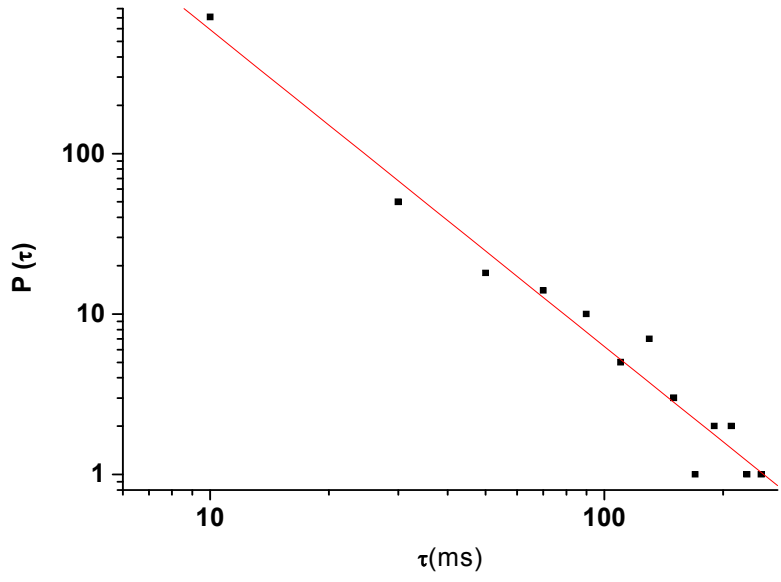

Figure . 3a

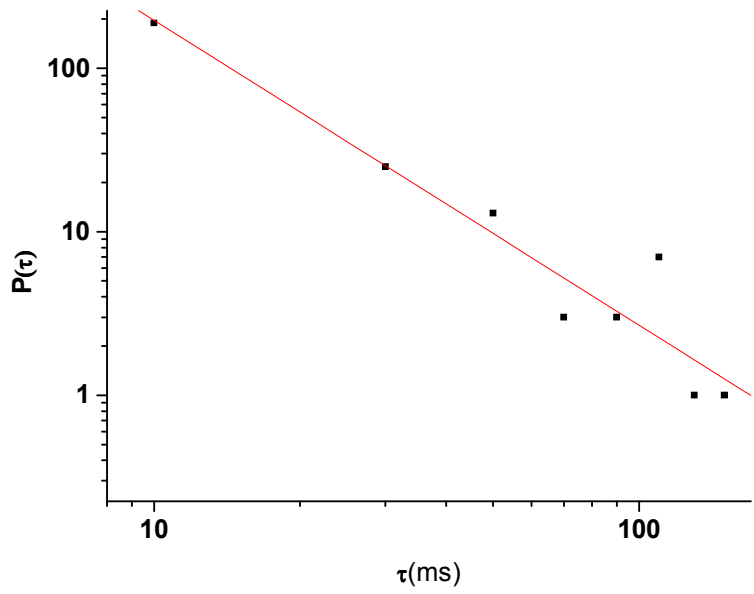

Figure. 3b

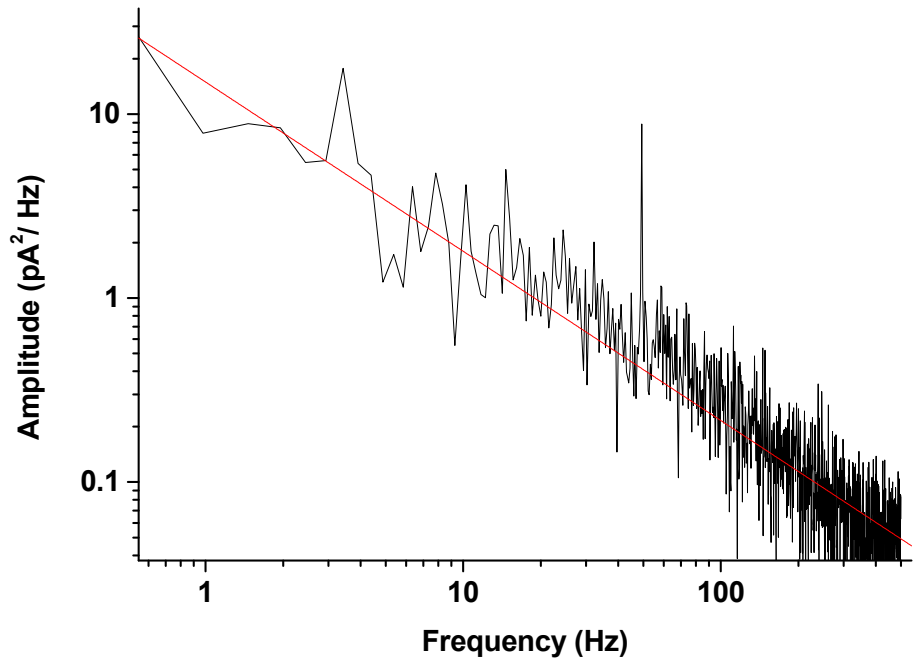

Figure. 4